\documentclass[preprint]{aastex}
\usepackage{CJK}

\renewcommand{\vec}[1]{\mbox{\boldmath{$#1$}}}

\newcommand{\eqref}[1]{(\ref{#1})}
\newcommand{\del}{\vec{\nabla}}
\newcommand{\pder}[2]{\frac{\partial #1}{\partial #2}}
\newcommand{\tder}[2]{\frac{d#1}{d#2}}

\newcommand{\sh}{$H$}
\newcommand{\bdim}[3]{#1\sh$\times$#2\sh$\times$#3\sh}
\newcommand{\rhopmax}{\rho_{p,\mathrm{max}}}

\shorttitle{Planetesimal Formation by Streaming Instability}
\shortauthors{Yang \& Johansen}

\begin{document}
\begin{CJK*}{UTF8}{bkai}

\title{On the Feeding Zone of Planetesimal Formation\\
       by the Streaming Instability}

\author{Chao-Chin Yang (楊朝欽) and Anders Johansen}
\affil{Lund Observatory, Department of Astronomy and Theoretical Physics, Lund University,\\
    Box~43, SE-22100 Lund, Sweden}
\email{ccyang@astro.lu.se, anders@astro.lu.se}

\begin{abstract}
The streaming instability is a promising mechanism to overcome the barriers in direct dust growth and lead to the formation of planetesimals.  Most previous studies of the streaming instability, however, were focused on a local region of a protoplanetary disk with a limited simulation domain such that only one filamentary concentration of solids has been observed.  The characteristic separation between filaments is therefore not known.  To address this, we conduct the largest-scale simulations of the streaming instability to date, with computational domains up to 1.6 gas scale heights both horizontally and vertically.  The large dynamical range allows the effect of vertical gas stratification to become prominent.  We observe more frequent merging and splitting of filaments in simulation boxes of high vertical extent.  We find multiple filamentary concentrations of solids with an average separation of about 0.2 local gas scale heights, much higher than the most unstable wavelength from linear stability analysis.  This measures the characteristic separation of planetesimal forming events driven by the streaming instability and thus the initial feeding zone of planetesimals.
\end{abstract}


\section{INTRODUCTION}

It is a long standing problem in the theory of planet formation how to grow centimeter/meter-sized solid objects into kilometer-sized planetesimals in a gaseous protoplanetary disk.  Such large particles are prone to bouncing as well as fragmentation under mutual collisions, making growth by coagulation inefficient \citep{ZO10}.  Their growth is still possible by mass transfer, when a small impactor hits a much larger target \citep{WB12a,WB12b,GM13}.  However, the subsequent growth is slow and requires artificial injection of centimeter-sized seeds among the millimeter-sized particles stuck at the bouncing barrier.  Furthermore, macroscopic solid particles with friction times comparable to the orbital period lose angular momentum to the gas, causing orbital decay in as little as 100~orbits, due to the drag of the head wind of the slower moving gas which is slightly pressure supported \citep{AHN76,sW77}.  Sufficiently dense protoplanetary disks may trigger gravitational instability in the dust mid-plane layer to form planetesimals locally in a direct gravitational collapse of solid materials \citep{GW73}, but turbulent diffusion prevents solids from sedimenting and reaching critical density in the mid-plane of the disk \citep{sW80}, unless the metallicity is significantly enhanced \citep{mS98,YS02,eC08}.

One of the most promising mechanisms to overcome these barriers is through the streaming instability.  It was discovered analytically by \citet{YG05}, who were inspired by an earlier simplified mid-plane layer model of \citet{GP00}, and numerically confirmed by \citet{YJ07} and \citet{JY07}.  The streaming instability arises in the mutual friction between the gas and the solids, with which the radial drift of the solids is reduced with increased mass loading.  Since the speed reduction is proportional to the solid density, a local concentration of solids migrates slower than isolated particles and accumulates the faster-migrating upstream materials, further reducing the drift speed of the over-dense region.  With this positive feedback loop, it has been shown that the local solid density can be enhanced by three orders of magnitude above the mean gas density in the mid-plane, triggering gravitational collapse to produce Ceres-sized planetesimals \citep{JO07} and smaller \citep{JYL12}, depending on the local column density of solids.

An ever increasing understanding of the streaming instability has been obtained during the past few years.  It operates in both laminar disks \citep{JYM09,BS10b} and in background turbulence driven by the magneto-rotational instability \citep{JO07,BT09,KFI12}.  There exists a critical solid-to-gas ratio above which strong clumping of solids occurs \citep{JYM09,BS10b}, and the ratio depends on the radial pressure gradient of the gas \citep{BS10a} and the size of the solid particles \citep[Carrera, Johansen, \& Davies, in preparation; see discussion in][]{JB14}.  Besides these studies, which are predominantly in the framework of the local-shearing-box approximation, it has been confirmed that the streaming instability operates in a global unstratified model, with results which are consistent with those found in the local approach \citep{KH13}.

It remains unclear, however, whether enough dynamical range for the nonlinear evolution of the streaming instability has been captured.  All of the previous works either ignored vertical stratification of the gas or did not cover enough vertical range so that the stratification became conspicuous.  Furthermore, in simulations including sedimentation of the particles, only one predominantly axisymmetric filamentary structure has been observed.  In this paper, we simulate the nonlinear evolution of the streaming instability in large computational domains, up to a factor of eight times larger than in previous works.  Indeed, we find that vertical stratification of the gas significantly influences the nonlinear evolution of the streaming instability.  We also capture multiple radial concentrations of the solid particles.  The former should serve as a steppingstone to establishing how the streaming instability interacts with the bulk of the gas in a more realistic protoplanetary disk model, while the latter helps characterize the typical separation between over-dense filaments and thus the feeding zone of planetesimal formation by the streaming instability.

This paper is organized as follows.  In Section~\ref{S:eom}, we describe the system of equations in our model and the numerical method we use for solving them.  In Section~\ref{S:results}, we measure the properties of the particle mid-plane layer where the streaming instability operates and describe how the characteristics of the mid-plane layer depend on the dimensions of the simulation box as well as the resolution.  In Section~\ref{S:impn}, we discuss the implications for planetesimal formation in general, and particularly the asteroid belts.  We conclude in Section~\ref{S:conc}.

\section{EQUATIONS OF MOTION} \label{S:eom}

We continue to employ the classic local-shearing-box approximation as in many previous studies of the streaming instability \citep{GL65,BN95,HGB95}.  A rectangular box co-rotating with the local Keplerian velocity at its center is considered.  The orientation of the box in the $x$, $y$, and $z$ directions are always radial, azimuthal, and vertical, respectively.  It is also assumed that the size of the box is small compared to its distance to the central star.  Under these assumptions, the equations of motion can be linearized in terms of the position relative to the center of the box and the velocity relative to the local Keplerian flow.  In the following, we briefly describe the equations of motion for the fluid gas and the solid particles as well as the numerical method we use to solve them.

\subsection{Gas}

For simplicity, we only consider a non-magnetized gas disk such that the magneto-rotational instability \citep[e.g.,][]{BH91} is not operating and thus the basic state of the gas is laminar.  We also assume an isothermal equation of state, which remains a good approximation given that the flow is strongly subsonic and hence that any small temperature increase is radiated away efficiently.

The equations of motion for the gas then become
\begin{eqnarray} \label{E:eom}
  \pder{\rho_g}{t}
  + u_{0,y}\pder{\rho_g}{y}
  + \del\cdot(\rho_g\vec{u})
  &=& 0,\label{E:continuity}\\
  \pder{\vec{u}}{t}
  + u_{0,y}\pder{\vec{u}}{y}
  + \vec{u}\cdot\del\vec{u}
  &=& -c_s^2\del\ln\rho_g
  + \left(2\Omega_K u_y\hat{\vec{x}}
          - \frac{1}{2}\Omega_K u_x\hat{\vec{y}}
          - \Omega_K^2 z\hat{\vec{z}}\right)\nonumber\\ &&
  + 2\Omega_K\Delta v\vec{\hat{x}}
  + \frac{\rho_p}{\rho_g}
    \frac{\vec{v} - \vec{u}}{t_s}.\label{E:momentum}
\end{eqnarray}
The dependent variables we solve for are the gas density $\rho_g$ and the gas velocity $\vec{u}$ relative to the background shear flow $\vec{u}_0 = -(3/2)\Omega_K x \vec{\hat{y}}$, with $\Omega_K$ being the local Keplerian angular frequency.  The parameter $c_s$ is the speed of sound, being constant.  The first two terms in the parenthesis of Equation~\eqref{E:momentum} result from the combination of the stellar radial gravity and the centrifugal and the Coriolis forces, while the last term accounts for the linearized stellar vertical gravity.  The next-to-last term on the right-hand side of Equation~\eqref{E:momentum} resembles the centrifugal support due to the large-scale radial pressure gradient in the gas disk such that the azimuthal speed of the gas is reduced by approximately
\begin{equation}
  \Delta v = \frac{1}{2}c_s\left(\frac{c_s}{v_K}\right)
             \left(-\pder{\ln p}{\ln R}\right),
\end{equation}
in which $v_K$ is the Keplerian velocity, $p$ is the gas pressure, and $R$ is the radial distance to the central star.\footnote{The ratio $\Delta v / c_s$ is equal to the dimensionless parameter $\Pi$ defined by \citet{BS10b}, and $\Delta v = \eta v_K$ with the dimensionless parameter $\eta$ defined by \citet{NSH86} \citep[see also][]{YG05}.}  In this work, we set $\Delta v / c_s = 0.05$, which is a typical value in the inner regions of a minimum-mass-solar-nebular disk model \citep{cH81,BS10b}.  The gas experiences the frictional drag from the solid particles through the last term in Equation~\eqref{E:momentum}, where $\rho_p$ is the volume density of solids, $\vec{v}$ is the local velocity of particles relative to the background shear, and $t_s$ is the stopping time for the particles (see Section~\ref{SS:particles}).  The factor $\rho_p / \rho_g$ stems from the conservation of linear momentum in the friction between the gas and the particles.

The primary objective of this work is to simulate the streaming instability in relatively large boxes.  In this regard, the vertical density stratification of the gas becomes significant, and preserving this stratification numerically is to our advantage here.  We define
\begin{equation}
  \rho_g(x,y,z,t) \equiv \rho_{g,0}(z)\left[1 + \xi(x,y,z,t)\right],
\end{equation}
where
\begin{equation} \label{E:denstrat}
  \rho_{g,0}(z) = \rho_0 \exp\left(-\frac{z^2}{2H^2}\right)
\end{equation}
is a constant background density stratification, which is set by the balance between stellar vertical gravity and gas pressure, and the gas scale height is thus $H = c_s / \Omega_K$.  The arbitrary constant $\rho_0$ is the mid-plane density of this equilibrium stratification, which depends on the location of the shearing box in the protoplanetary disk.  Note that positive densities imply $\xi > -1$.  The equations of motion formulated in $\xi$ now read
\begin{eqnarray}
  \pder{\xi}{t}
  + u_{0,y}\pder{\xi}{y}
  + \vec{u}\cdot\del\xi
  + (1 + \xi)\del\cdot\vec{u}
  &=& \frac{z u_z}{H^2}(1 + \xi),\label{E:strat-con}\\
  \pder{\vec{u}}{t}
  + u_{0,y}\pder{\vec{u}}{y}
  + \vec{u}\cdot\del\vec{u}
  &=& -c_s^2\del\ln(1 + \xi)
  + \left(2\Omega_K u_y\hat{\vec{x}}
        - \frac{1}{2}\Omega_K u_x\hat{\vec{y}}\right)\nonumber\\&&
  + 2\Omega_K\Delta v\vec{\hat{x}}
  + \frac{\rho_p}{\rho_g}
    \frac{\vec{v} - \vec{u}}{t_s}.\label{E:strat-mom}
\end{eqnarray}
The stellar vertical gravity exactly cancels the pressure gradient of the equilibrium density stratification in the momentum Equation~\eqref{E:strat-mom}, while a source term appears in the continuity Equation~\eqref{E:strat-con} to account for the imbalance in vertical mass flux due to the stratification.  With this formulation, we solve for the dimensionless variable $\xi$ instead of $\rho_g$.

We use the usual sheared periodic boundary conditions \citep[e.g.,][]{HGB95}, where $f(x,y,z) = f(x + L_x, y - (3/2)\Omega_K L_x t, z)$ in which $f$ is any field in question and $L_x$ is the $x$-dimension of the computational domain.  We adopt periodic boundary conditions for the vertical direction.  Note that in the formulation of Equations~\eqref{E:strat-con} and~\eqref{E:strat-mom}, a periodic $\xi$ in the vertical direction does not introduce discontinuity in either density or pressure gradient \citep[c.f.,][]{DSP10}.  We set $\xi = 0$ and $\vec{u} = 0$ at $t = 0$ as our initial conditions.

\subsection{Particles} \label{SS:particles}

Instead of treating particles as pressureless fluid, we consider the motion of each individual particle according to
\begin{eqnarray}
  \tder{\vec{x}_p}{t}
  &=& -\frac{3}{2}\Omega_K x_p \vec{\hat{y}}
  + \vec{v},\label{E:part_vel}\\
  \tder{\vec{v}}{t}
  &=& \left(2\Omega_K v_y\hat{\vec{x}}
  - \frac{1}{2}\Omega_K v_x\hat{\vec{y}}
  - \Omega_K^2 z_p \vec{\hat{z}}\right)
  + \frac{\vec{u} - \vec{v}}{t_s},\label{E:part_acc}
\end{eqnarray}
where $\vec{x}_p = (x_p, y_p, z_p)$ is the position of the particle relative to the center of the box and $\vec{v} = (v_x, v_y, v_z)$ is the relative velocity of the particle with respect to the Keplerian shear as defined above.  Equation~\eqref{E:part_vel} is the total velocity of the particle while Equation~\eqref{E:part_acc} is the acceleration of the particle with the contributions parallel to those for the gas except the pressure gradient (see Equation~\eqref{E:momentum}).

The stopping time $t_s$ in Equations~\eqref{E:momentum}, \eqref{E:strat-mom}, and \eqref{E:part_acc} is the damping time for the relative speed between gas and each solid particle due to their mutual viscous drag.  It is often expressed in terms of the dimensionless parameter $\tau_s \equiv \Omega_K t_s$, which is a measure of the Stokes number for the particles.  In this work, we set $\tau_s = \pi / 10 \simeq 0.314$, for which the radius of the particles is about 0.7~m at 1~AU or about 4~mm at 30~AU in the minimum-mass solar nebula \citep{JO07,BS10b,JB14}.

It is impractical to simulate all millimeter-to-meter-sized solid particles even with a computational box as small as one of 0.2$H$ on an edge.  Instead, we consider super-particles, each of which represents a swarm of real, identical particles.  The mass of each super-particle is the total mass of the constituent particles, while the damping time for the super-particle remains the same as that of the individual members.  It has been demonstrated that this approach is numerically convergent when on average more than one super-particles per grid cell exist in the sedimented mid-plane layer \citep{YJ07,BS10c}.

The mass of each super-particle $m_p$ is determined by the initial solid-to-gas ratio, $Z \equiv \Sigma_{p,0} / \Sigma_{g,0}$, of the medium, where $\Sigma_{p,0}$ and $\Sigma_{g,0}$ are the initial (uniform) column densities of the particles and the gas, respectively, integrated over the full vertical extent of the disk.  From Equation~\eqref{E:denstrat}, $\Sigma_g = \sqrt{2\pi}H\rho_0$.  Since virtually all particles we consider settle within a distance much less than $H$ of the mid-plane of the disk, $\Sigma_p$ is well approximated by $m_p N_p / \left(L_x L_y\right)$, where $N_p$ is the total number of particles used in a simulation, and $L_x$ and $L_y$ are the sizes of the computational domain in the $x$ and the $y$ directions, respectively.  In this work, we set $Z = 0.02$, which is just above the critical solid-to-mass ratio required to trigger strong clumping of particles by the streaming instability \citep{JYM09}.  In addition, we use as many particles as the total number of grid points in each simulation, i.e., $N_p = N_x N_y N_z$, where $N_x$, $N_y$, and $N_z$ are the number of grid points in the $x$, $y$, and $z$ directions.  Therefore, the average number of particles per cell near the mid-plane is roughly $N_z / N_\mathrm{mid}$ after sedimentation, where $N_\mathrm{mid}$ is the number of vertical grid cells resolving the particle scale height.

As our initial conditions, we use a uniform distribution to randomly place the particles throughout the computational domain while setting $\vec{v} = 0$.  The noise inherent in the initial positions of the particles serve as the seed for the ensuing growth of the streaming instability.  The boundary conditions for the particles are such that when a particle crosses a boundary plane, it reemerges in the opposite plane with sheared periodic positional mapping as the gas while preserving its relative velocity $\vec{v}$.

\subsection{Numerical Method}

To solve the system of Equations~\eqref{E:strat-con}, \eqref{E:strat-mom}, \eqref{E:part_vel}, and~\eqref{E:part_acc}, we use the Pencil Code\footnote{The Pencil Code is publicly available at \texttt{https://code.google.com/p/pencil-code/}.}, a cache-efficient, parallelized magnetohydrodynamical code founded by \citet{BD02}.  For the (magneto-)hydrodynamics, the code employs sixth-order finite differences in space while integrating the system of equations in time by third-order Runge-Kutta steps.  For the particle dynamics, the position and velocity of each individual particle is evolved simultaneously with the Runge-Kutta time steps for the fluid.  The interactions between the fluid and the particles, i.e., the frictional drag, are computed via the standard particle-mesh method of triangular-shaped clouds, which ensures conservation of total momentum \citep{YJ07,JO07}.  The particle-block-decomposition algorithm implemented by \citet{JKH11} is used in order to achieve better load balance among processors.

We have implemented in the Pencil Code the new formalism for balanced stratification of gas density, Equations~\eqref{E:strat-con} and~\eqref{E:strat-mom}.  We find this formalism is in general more numerically stable than evolving the system of Equations~\eqref{E:continuity} and~\eqref{E:momentum} in the sense that much less artificial diffusion is required to maintain hydrostatic equilibrium against perturbations.  This is especially true when the vertical extent of the computational domain is greater than a few disk scale heights.  Furthermore, this formalism relieves the necessity of implementing special boundary conditions in order to incorporate the background density stratification.  Finally, we find that the total mass in the system remains well conserved with this new formalism, in spite that the continuity equation (Equation~\eqref{E:strat-con}) is not written in conservative form.

In this work, we also increase the time-step constraint limited by the drag force calculation from one-fifth of the decay time constant adopted in previous works to one time constant.  This amounts to a possible over-damping in relative velocity between gas and particles by a maximum relative error of about $9\%$.  We find that this relaxation does not noticeably alter the results while allowing us a five-time speed up for otherwise the same simulation.

We systematically adjust the dimensions of the simulation box and investigate the difference in the results.  The horizontal sizes $L_x = L_y$ span from 0.2$H$ up to 1.6$H$, while the vertical size $L_z$ is up to 1.6$H$ with $L_z \leq L_x$.  The maximum number of grid cells in each dimension we have explored is 256, which translates to a resolution of 160~$H^{-1}$ for our largest \bdim{1.6}{1.6}{1.6} box.

\section{SIMULATION RESULTS} \label{S:results}

In the following subsections, we report several properties of the resulting distribution of particles from our simulations and discuss their dependence on box dimensions and resolution.

\subsection{Particle Scale Height} \label{SS:psh}

\begin{figure}[htbp]
\begin{center}
\plotone{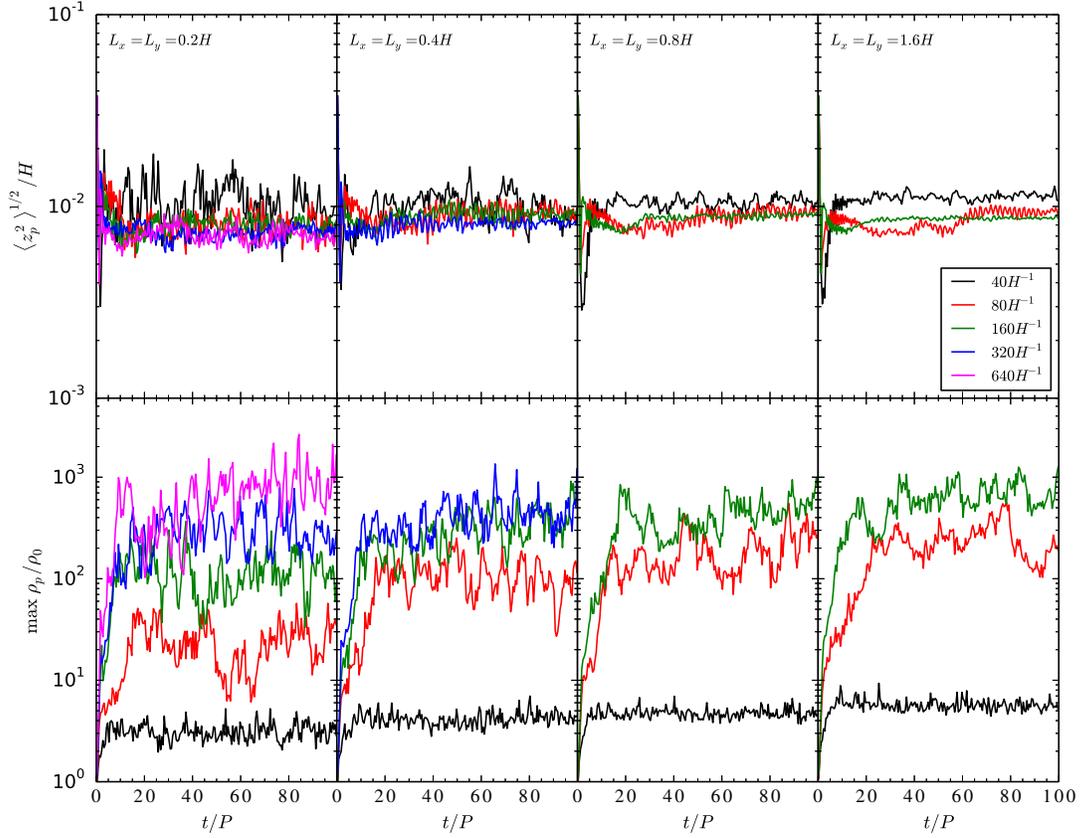}
\caption{Average particle scale height (top row) and maximum local particle density (bottom row) as a function of time for various resolutions as well as horizontal sizes of the computational domain.  Different columns correspond to different horizontal sizes, while different lines represent different resolutions.  The vertical size of the computational domain is $L_z = 0.2~H$.}
\label{F:plres}
\end{center}
\end{figure}

The top row of Figure~\ref{F:plres} shows the particle scale height as a function of time for boxes with various resolutions and horizontal sizes but a constant vertical size of $L_z = 0.2H$.  Since the particles are initially uniformly distributed, the initial scale height is virtually infinite.  Nevertheless, the particles quickly settles down toward the mid-plane within $t = 2P$ due to their vertical motion and gas drag, where $P = 2\pi / \Omega_K$ is the orbital period.  At this point, the particles have concentrated near the mid-plane to the extent that the gas experiences enough perturbation from the particles to become turbulent.  The random motion of the gas-dragged particles becomes dominant and puffs the particle layer back up.  Within another time interval of $\sim$1$P$, the particle scale height reaches its final, roughly constant value, which lasts till the end of the simulations at $t = 100P$.

Little variation in the equilibrium particle scale height is observed between boxes of different resolutions and/or different horizontal sizes.  Only the boxes with a resolution of 40$H^{-1}$, the lowest we have investigated in this work, show a slightly larger particle scale height.  The equilibrium scale height is on the order of $\sim$10$^{-2}H$, which is resolved beyond a resolution of 320$H^{-1}$.  Nevertheless, it is not clear if resolving the particle layer is critical in the saturated state of the streaming instability except perhaps in predicting the correct peak local particle density (see below).

The top row of Figure~\ref{F:pllz} also shows the particle scale height as a function of time, but for boxes of various horizontal and vertical dimensions at a fixed resolution of 160$H^{-1}$.  Similarly to the case of $L_z = 0.2H$ discussed above, the horizontal size has little effect on the particle scale height for other vertical dimensions.  On the other hand, we see a factor of close to two increase in the particle scale height from a box of $L_z = 0.2H$ to that of $L_z = 0.4H$.  Boxes of $L_z \gtrsim 0.4H$ have a consistent equilibrium value.  However, it appears that the larger the vertical size of the box, the longer time scale is required to reach the equilibrium, which might be an effect of the more stochastic particle clumping for boxes with larger $L_z$ (see Section~\ref{SS:sigptx}).
\begin{figure}[htbp]
\begin{center}
\plotone{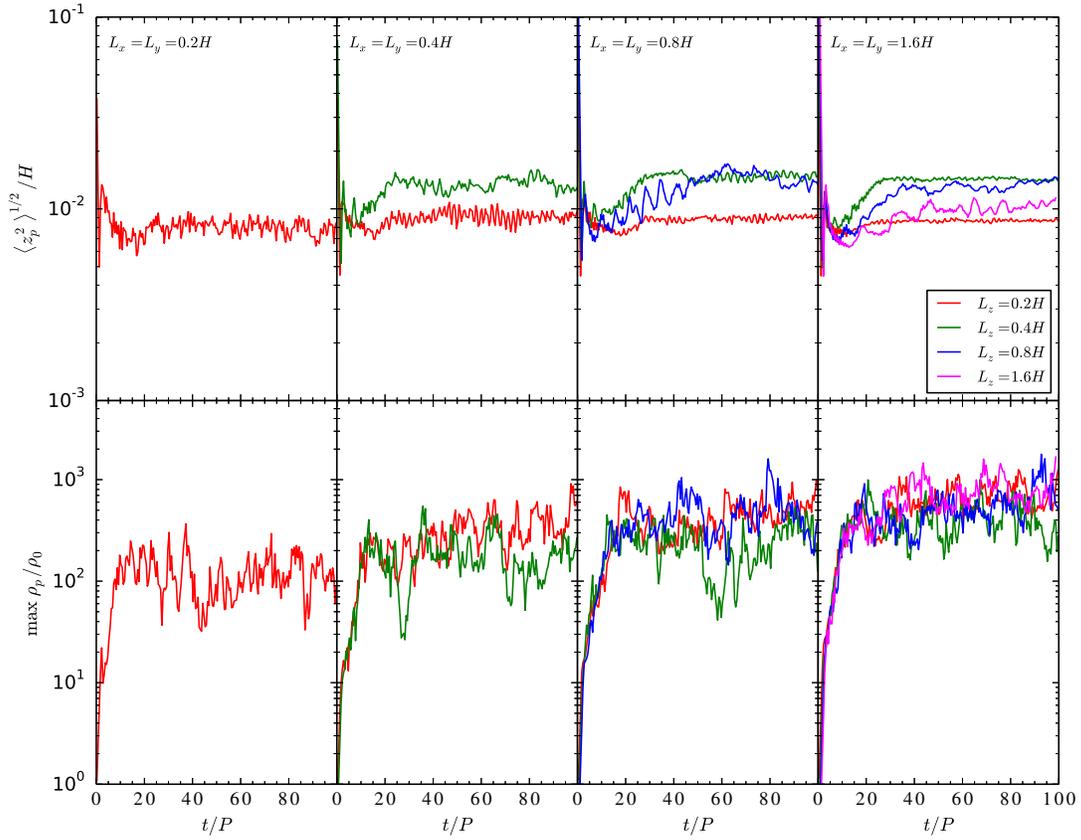}
\caption{Average particle scale height (top row) and maximum local particle density (bottom row) as a function of time for various sizes of the computational domain.  Different columns correspond to different horizontal sizes, while different lines represent different vertical sizes.  The resolution is at 160~points per gas scale height.}
\label{F:pllz}
\end{center}
\end{figure}

\subsection{Maximum Particle Density} \label{SS:mpd}

The bottom rows of Figures~\ref{F:plres} and~\ref{F:pllz} show the corresponding maximum particle density $\rhopmax$ as a function of time for various box dimensions and resolutions.  Due to the vertical sedimentation of the particles, the particle density coherently increases within $t = 2P$ to above the initial gas density $\rho_0$ in the mid-plane.  As can be seen in Figure~\ref{F:plres}, the particle density for the boxes with a resolution of 40$H^{-1}$ then remains at a constant low level of a few $\rho_0$ and no significant particle concentration occurs.  On the other hand, appreciable particle concentration driven by the streaming instability starts to appear and drives $\rhopmax$ further up for all boxes with a resolution of $\gtrsim$80$H^{-1}$, before it reaches a roughly constant state at $t \sim 20P$.

In general, the higher the resolution, the larger the maximum particle density $\rhopmax$ results, which has been reported in previous studies \citep{JO07,BS10c,JYL12}.  Here, we also find the higher the resolution, the smaller the increase in the final level of $\rhopmax$.  This indicates the numerical convergence with resolution in the saturated stage of the streaming instability.  And Figure~\ref{F:plres} hints that a resolution of $\sim$160--320$H^{-1}$ might already give a converged result in $\rhopmax$, at least in the case of this work.  A level of $\rhopmax$ close to 10$^3\rho_0$ has been reached.

As shown in Figure~\ref{F:pllz}, we find little variation in the maximum particle density $\rhopmax$ in the saturated stage of the streaming instability with respect to the vertical box size $L_z$.  On the other hand, there exists a slight increase of a factor of a few in $\rhopmax$ from horizontal box size $L_x = L_y = 0.2H$ to $L_x = L_y = 0.4H$, while the results are consistent for all boxes with $L_x = L_y \gtrsim 0.4H$.

Combining with the similar behavior of the particle scale height discussed in Section~\ref{SS:psh}, this suggests that simulation boxes with either horizontal extent $L_x = L_y = 0.2H$ or vertical extent $L_z = 0.2H$ are insufficient to capture all necessary scales perturbed by the streaming instability in its nonlinear saturation stage.  More evidence on this is presented in Section~\ref{SS:sigptx}.

\subsection{Characteristics of the Particle Radial Concentration} \label{SS:sigptx}

The streaming instability predominantly concentrates sedimented particles radially into filamentary structures, extended in the azimuthal direction \citep{JY07,BS10b,KH13}.  Therefore, the column density of particles, while averaged over the azimuthal dimension of the simulation box, well describes the time evolution of the particle layer driven by the streaming instability.  Particular interest here is to investigate the dependence of these particle radial concentrations on the box dimensions as well as the resolution.

\begin{figure}[htbp]
\epsscale{0.85}
\plotone{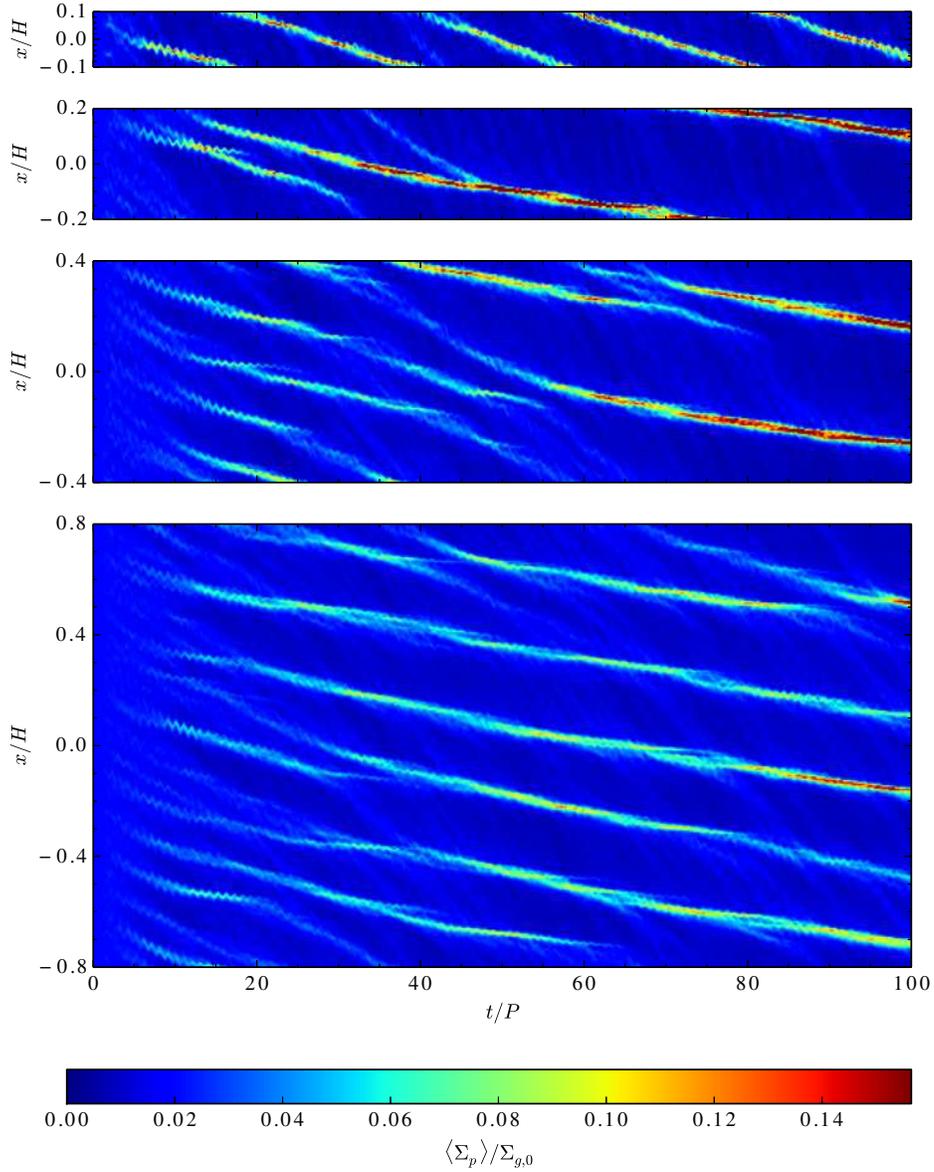}
\epsscale{1}
\caption{Particle column density averaged over azimuth $\langle\Sigma_p\rangle$ as a function of time $t$ and radial location $x$ for simulation boxes of various horizontal sizes. The horizontal box sizes are, from top to bottom, $L_x = L_y = 0.2$~\sh, 0.4~\sh, 0.8~\sh\ and 1.6~\sh, respectively.  Fixed are the vertical box size at $L_z = 0.2~H$ and the resolution at 160 $H^{-1}$.}
\label{F:sigph}
\end{figure}

Figure~\ref{F:sigph} shows the averaged column density of particles $\langle\Sigma_p\rangle$ as a function of time and radial position for the simulation boxes with the same resolution (160$H^{-1}$) and vertical extent (0.2\sh) but varying horizontal dimensions (from 0.2$H$ to 1.6$H$).  The \bdim{0.2}{0.2}{0.2} box shows only one major filament of solids, which is consistent with previous works \citep[e.g.,][]{JYM09}.  The \bdim{0.4}{0.4}{0.2} box, however, generates two filaments initially, but one of them disperses and later merges into the other, forming one dense filament.  More interestingly, the \bdim{0.8}{0.8}{0.2} box and the \bdim{1.6}{1.6}{0.2} box demonstrate much richer dynamics driven by the streaming instability, with multiple filaments of solids forming, dispersing, splitting, and merging, in drastic contrast to one single dominant filament for smaller simulation boxes.  For the \bdim{1.6}{1.6}{0.2} box, about five or six dense filaments coexist at any given time.  Noticeable is that their separation remains quite regular for a long period of time, due to roughly the same radial drift speed of the filaments.  Since the radial drift speed is determined by the particle density \citep{NSH86,YG05}, this in turn implies that the filaments have roughly the same column density, as is also seen in the figure.

\begin{figure}[htbp]
\epsscale{0.85}
\plotone{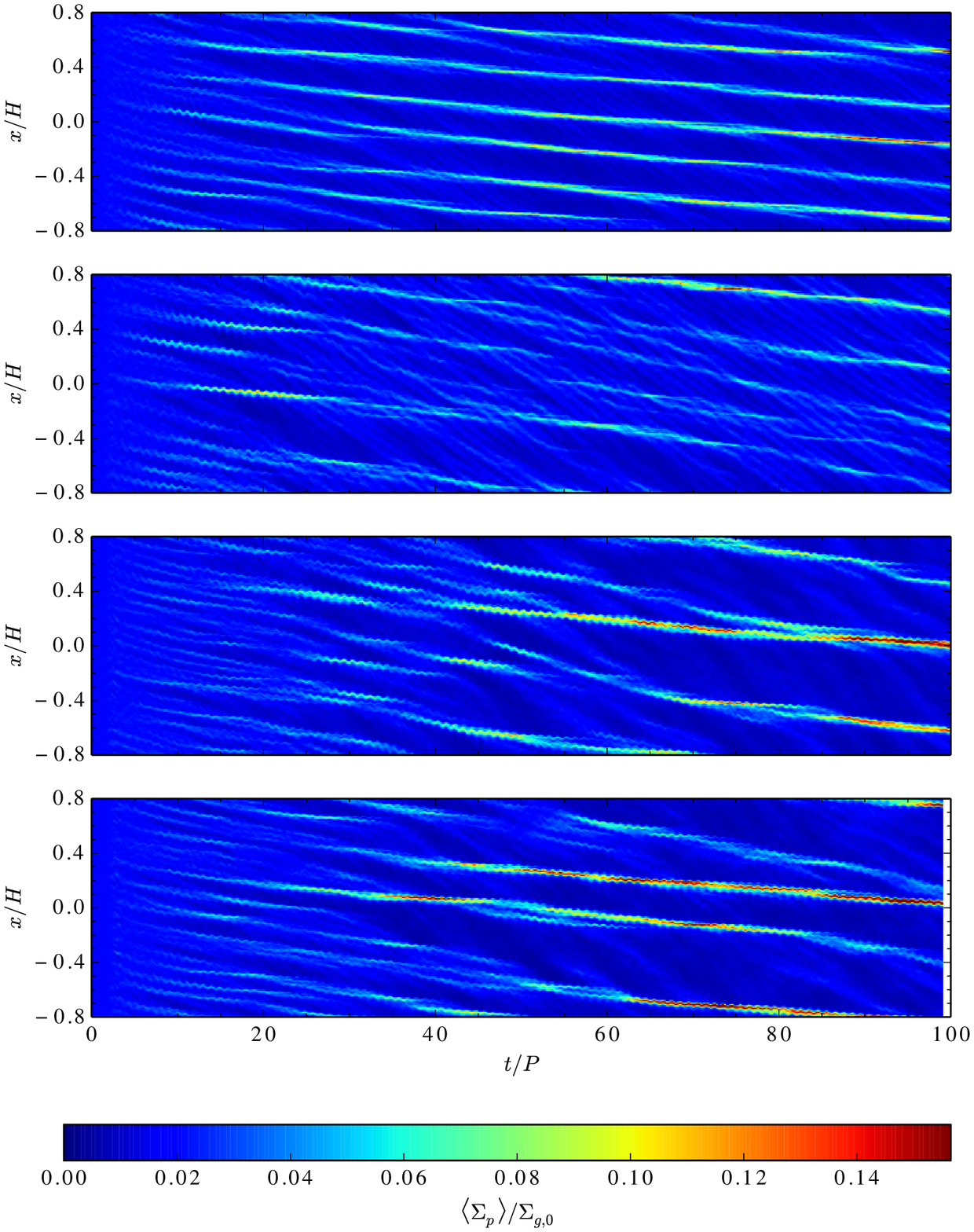}
\epsscale{1}
\caption{Particle column density averaged over azimuth $\langle\Sigma_p\rangle$ as a function of time $t$ and radial location $x$ for simulation boxes of various vertical sizes. The vertical box sizes are, from top to bottom, $L_z=$ 0.2~\sh, 0.4~\sh, 0.8~\sh\ and 1.6~\sh, respectively.  Fixed are the horizontal box size at $L_x = L_y = 1.6~H$ and the resolution at 160 $H^{-1}$.}
\label{F:sigpv}
\end{figure}

Figure~\ref{F:sigpv} compares $\langle\Sigma_p\rangle$ for the simulation boxes of various vertical sizes (from $L_z = 0.2H$ to 1.6$H$) but the same resolution (160$H^{-1}$) and horizontal dimensions ($L_x = L_y = 1.6H$).  Extending the vertical size of the box evidently introduces much more complexity in the evolution of the particle layer.  The densities of the particle filaments for the tall boxes have significantly larger variance than those for their short counterpart (the \bdim{1.6}{1.6}{0.2} box) such that their radial drift speeds noticeably differ.  This in turn makes the merging and splitting events of the filaments occur relatively more frequently.  Nevertheless, multiple filaments still remain at any given time for these tall boxes.

In order to make a more quantitative statement on their characteristics, we devise a simple algorithm to capture the particle filaments.  At any given time, we use a stencil of fixed physical length $w$ to scan through the radial position $x$, appending enough ghost cells near the two ends of the computational domain.  If the azimuthally averaged particle column density $\langle\Sigma_p\rangle$ at the center of a stencil is the maximum for all points in the stencil and is larger than a certain threshold, we define that a concentrated particle filament occurs at this location with its peak density the same as the maximum $\langle\Sigma_p\rangle$.  We choose $w = 0.05H$ and use for the threshold 5$\sigma$ Poisson noise in a uniform particle distribution, which is represented by $N_z$ particles for each column of cells in our simulations, i.e.,
\begin{equation}
    \max_{x_i - w/2 \leq x_j \leq x_i + w/2}\langle\Sigma_p\rangle_{x = x_j}
    = \langle\Sigma_p\rangle_{x = x_i}
    > \Sigma_{p,0}\left(1 + \frac{5}{\sqrt{N_z}}\right).
\end{equation}
Note that the minimum separation between adjacent filaments that can be resolved is then $w/2 = 0.025H$.  In our simulations, especially those with high resolutions ($\gtrsim$160$H^{-1}$), we do find that dense filaments undergo splitting and merging events with structures of length scale less than 0.025$H$.  However, the majority of these events are intermittent and the filaments recover their original states on a relatively short timescale.  The remaining ones do significantly change the properties of the filaments and are detectable by this simple algorithm.

\begin{figure}[htbp]
\begin{center}
\plotone{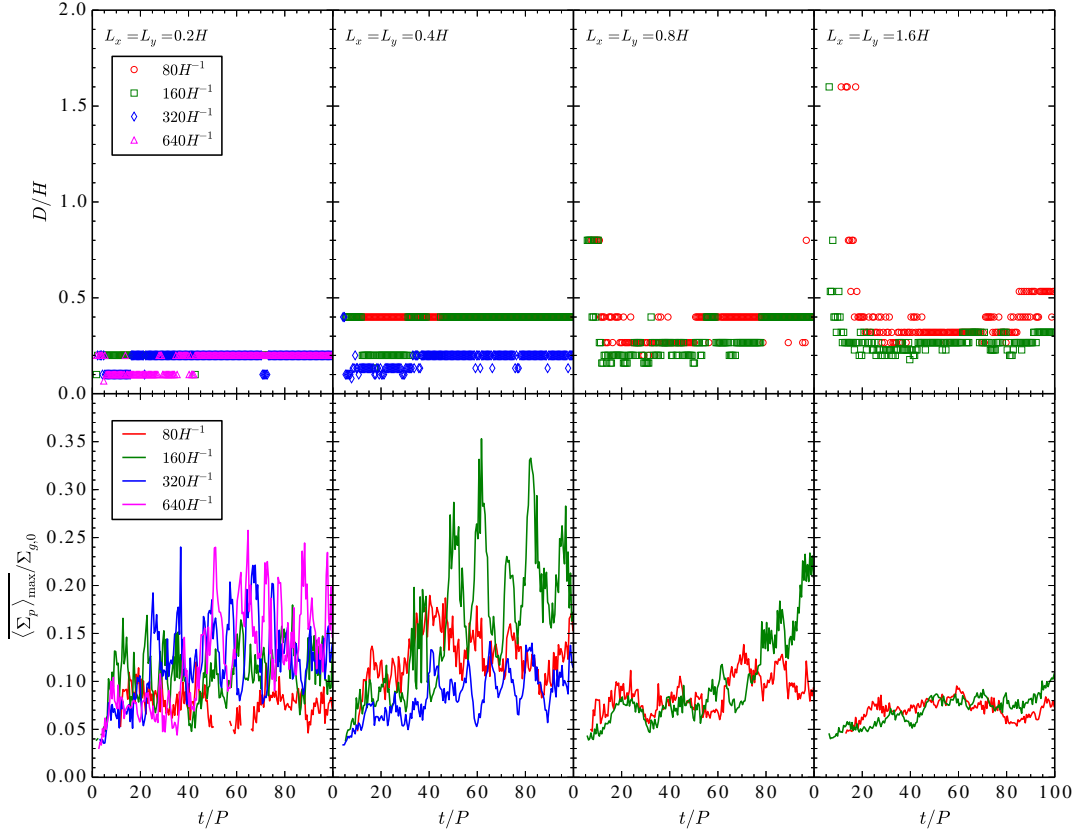}
\caption{Mean separation (top row) and mean peak column density (bottom row) of the azimuthal particle filaments as a function of time.  The box dimensions and resolutions correspond to those in Figure~\ref{F:plres}.}
\label{F:filares}
\end{center}
\end{figure}

Figure~\ref{F:filares} shows the mean separation $D$ and the mean peak averaged column density $\overline{\langle\Sigma_p\rangle_{\max}}$ of the particle filaments as a function of time for simulation boxes of various resolutions and horizontal dimensions but fixed vertical size $L_z = 0.2H$.  For boxes of the same dimensions, increasing the resolution generally results in more filaments being identified and thus smaller mean separation.  As can be seen in the figure, we find that the mean separation tends to converge with resolution towards  $D \sim 0.2H$, which is well above the radius of the stencil we use.  For smaller boxes with $L_x = L_y = 0.2H$ and 0.4$H$, the mean peak density $\overline{\langle\Sigma_p\rangle_{\max}}$ covers a wide range of values (between 0.05$\Sigma_{g,0}$ and 0.35$\Sigma_{g,0}$) with no obvious trend of convergence.  This is due to small-number statistics (only one or two filaments form in these cases) and indicates the stochastic nature in the formation and evolution of these filaments.  On the other hand, the larger boxes with $L_x = L_y = 0.8H$ and 1.6$H$ does show relatively consistent results between different resolutions and horizontal box sizes, where $\overline{\langle\Sigma_p\rangle_{\max}} \sim 0.08\Sigma_{g,0}$ .

\begin{figure}[htbp]
\begin{center}
\plotone{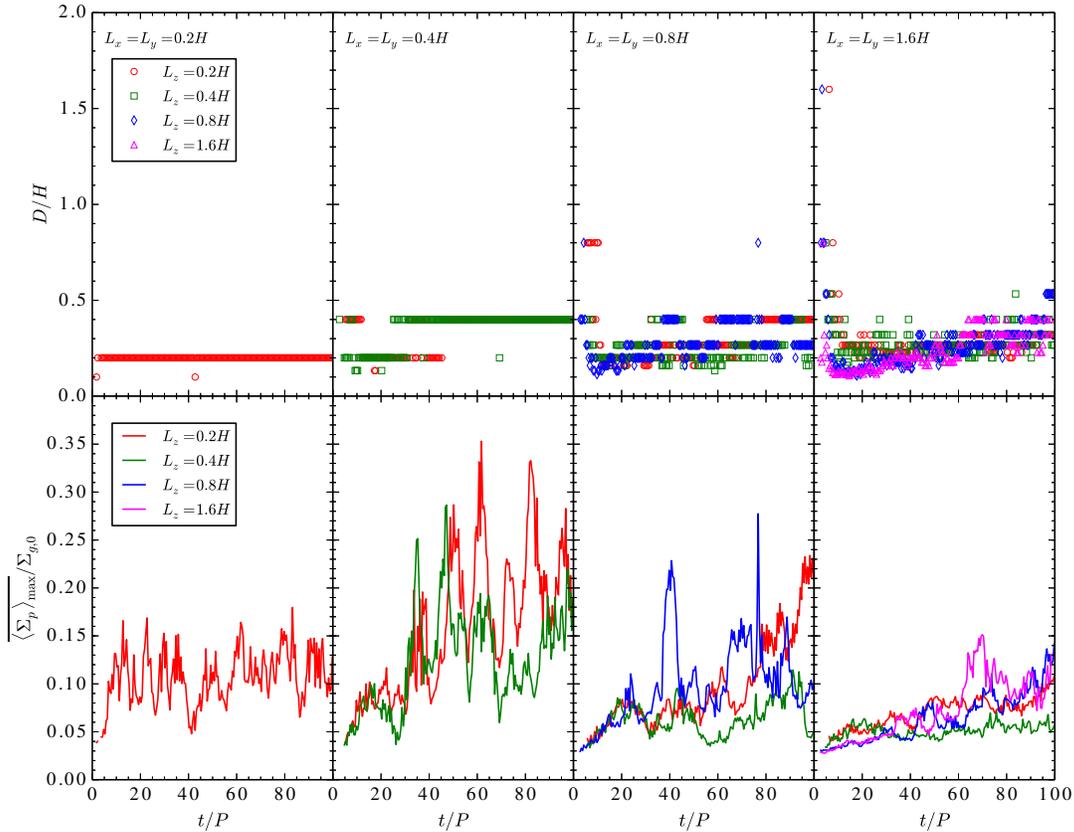}
\caption{Mean separation (top row) and mean peak column density (bottom row) of the azimuthal particle filaments as a function of time.  The box dimensions and resolutions correspond to those in Figure~\ref{F:pllz}.}
\label{F:filalz}
\end{center}
\end{figure}

Figure~\ref{F:filalz} further compares the mean separation $D$ and the mean peak averaged column density $\overline{\langle\Sigma_p\rangle_{\max}}$ for boxes of various vertical size $L_z$.  As discussed above, increasing $L_z$ makes the formation and evolution of the particle filaments even more stochastic.  This behavior is also manifest in $D$ and $\overline{\langle\Sigma_p\rangle_{\max}}$, where more time variation with more significant amplitude occurs for boxes of larger $L_z$.  This is yet another view of the more frequent formation, dispersal, splitting, and merging of the particle filaments along with larger variance in their densities for taller boxes.  Nevertheless, $D$ oscillates around $\sim$0.2$H$ for our largest boxes, indicating that the number of the major, persistent filaments is about the same over time.  These results demonstrate the importance of the vertical coverage of the gas disk to fully capture the dynamics driven by the streaming instability, even though the particle layer is thin compared to the gas scale height.

From Figures~\ref{F:filares} and~\ref{F:filalz}, we also notice that there exists a trend of increasing mean separation $D$ and the mean peak averaged column density $\overline{\langle\Sigma_p\rangle_{\max}}$ at late times for various cases.  This indicates the tendency for the particle filaments to merge and keep accreting surrounding materials.  However, we note that this phenomenon might not be relevant in reality, since from Figures~\ref{F:plres} and~\ref{F:pllz}, the local particle density, being the decisive factor to drive the ultimate gravitational collapse of the particles to form planetesimals, would have reached its maximum level already in the early saturated stage of the streaming instability.

\subsection{Gas-particle Correlation}
The results presented in Section~\ref{SS:sigptx} indicate that even though the particle layer is thin compared with the gas disk, the gas dynamics over at least one gas scale height cannot be ignored in the nonlinear stage of the streaming instability.  It appears that a significant fraction of the column of the gas mass is still required to better describe the interaction between the gas and the solids.  In this regard, we need to study the correlation between the gas and the solids and its dependence on the vertical dimension of the simulation box.

\begin{figure}[htbp]
\begin{center}
\plotone{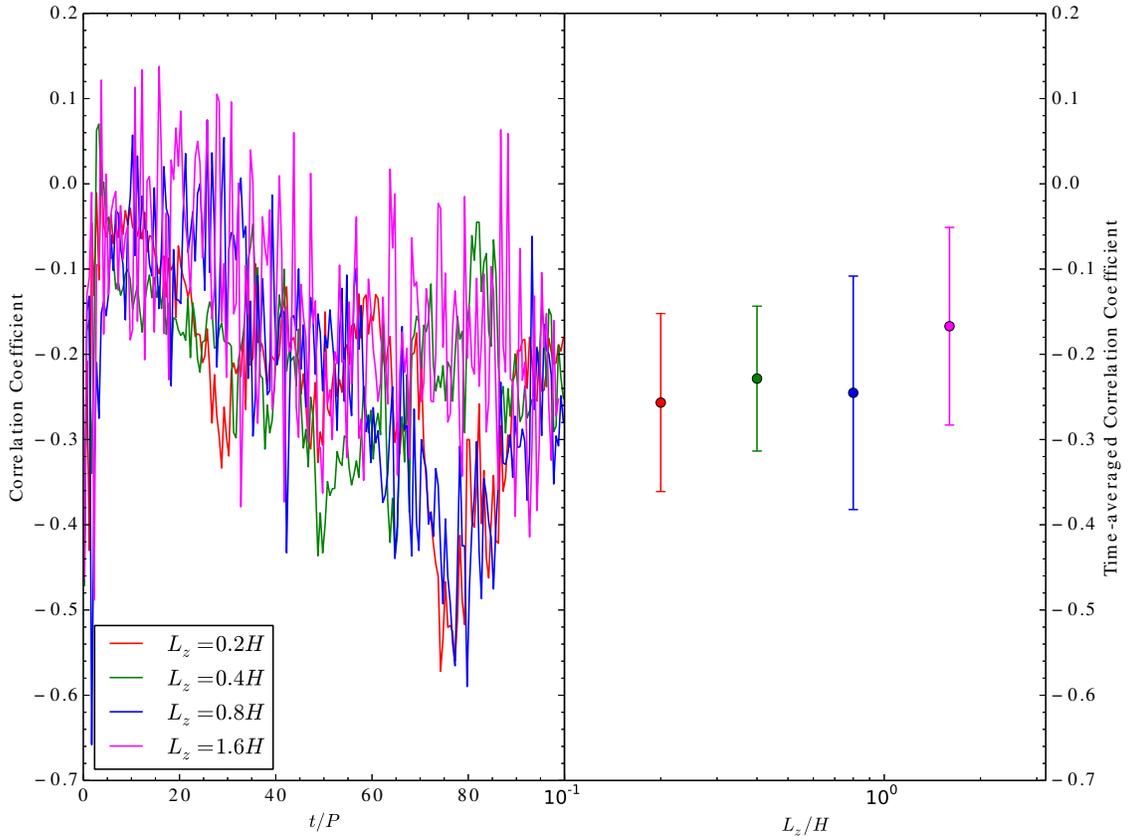}
\caption{The correlation coefficients between the $yz$-averages of the gas density deviation $\xi$ and of the solid density $\rho_p$ as a function of time for various vertical box sizes $L_z$ (\emph{left}) and the corresponding time averages over $t > 20P$ (\emph{right}).  The error bars represent one standard deviation.  The horizontal box dimensions are $L_x = L_y = 1.6H$ and the resolution is 160~$H^{-1}$.}
\label{F:gscorr}
\end{center}
\end{figure}

We plot in Figure~\ref{F:gscorr} the correlation coefficient between the $yz$-averages of the gas density deviation $\xi$ and of the solid density $\rho_p$ for various vertical box sizes.  The former is a proxy for the azimuthal average of the gas column density $\langle\Sigma_g\rangle$ and the latter is directly proportional to that of the solid column density $\langle\Sigma_p\rangle$, both of which are functions of the radial position $x$ and time $t$.  Since most of the solid particles are concentrated in azimuthal filamentary structures, the correlation coefficient represent the spatial correlation between the gas mass and the particle filaments.  As shown in the figure, although there exists significant time variation in the correlation coefficient, the gas distribution and the particle distribution in general \emph{anti}-correlate, with an average of about $-0.2$.  This implies that the gas tends to be entrapped in between the particle filaments and slightly enhance the pressure there.  We also see that this anti-correlation decreases noticeably with increasing vertical box size, indicating the lessening of the pressure buildup with increased vertical dynamical range.

Given these findings, we speculate that even though the perturbation in the gas due to the streaming instability is only at about 0.1\% level, it is enough for the gas to participate in regulating the dynamics of the particle filaments.  The slightly enhanced gas pressure between the filaments may help inhibit the filaments from approaching, leading to the exceptionally regular spacing and similar migration speeds as we see in Figure~\ref{F:sigph} for short simulation boxes.  With increased vertical dynamical range, on the other hand, the gas in the mid-plane gains freedom to escape vertically when the particle filaments tend to merge or split, relieving otherwise the pressure enhancement of the gas.  This may explain why there exists a noticeable decrease in the anti-correlation between the gas and the solid column densities with increasing vertical box size, as seen in Figure~\ref{F:gscorr}.  In any case, this further demonstrates the importance of the vertical dimension in the nonlinear evolution driven by the streaming instability.

\section{IMPLICATIONS FOR PLANETESIMAL FORMATION} \label{S:impn}

If planetesimals form from the dense particle filaments driven by the streaming instability, then the mean separation of the filaments delineates the mean radial separation of the new-born planetesimals.  It also characterizes the size of the feeding zone where the planetesimals keep accreting the surrounding materials.  Therefore, the variation in the composition of planetesimals may contain the information of the chemical inhomogeneity in their natal protoplanetary disk down to this length scale.

To put our measurement of the mean separation $D$ into perspective, we consider the minimum-mass-solar-nebula model \citep{cH81} as an example.  Then $D \sim 0.2H$ reads
\begin{equation}
  D \sim \left(0.02~\textrm{AU}\right)\left(\frac{R}{2.5~\textrm{AU}}\right)^{5/4}
      = \left(3\times10^6~\textrm{km}\right)\left(\frac{R}{2.5~\textrm{AU}}\right)^{5/4},
\end{equation}
where $R$ is the distance to the Sun.  And the total mass of the solid materials in an annulus of size $D$ is approximately
\begin{equation}
  M \simeq 2\pi Z\Sigma_g R D
      \sim \left(0.05~\textrm{M}_\earth\right)\left(\frac{Z}{0.01}\right)\left(\frac{R}{2.5~\textrm{AU}}\right)^{3/4}.
\end{equation}

Meteoritic classes show various degrees of aqueous alteration arising from the flow of liquid water inside the parent body.  The degree to which a planetesimal accretes ice is a measure of the distance to the snow line.  Among the chondrites, the enstatite chondrites EH and EL appear driest, followed by the ordinary chondrites, with the carbonaceous chondrites showing the highest degree of alteration \citep[e.g.,][]{SK03}.  The presence of these distinct classes could be a direct consequence of asteroid formation in discrete filaments resulting from the streaming instability, each of which probes an ice content set by the distance to the young Sun.

\section{CONCLUDING REMARKS} \label{S:conc}

We have performed a systematic study of how the nonlinear evolution of the streaming instability depends on the dimensions of the simulation box as well as on the resolution in a local, non-magnetized disk model.  In order to capture the vertical stratification of the gas, as well as cover more horizontal range, we have completed simulations with the largest computational domain of this kind to date, measuring 1.6 gas scale heights in each dimension, in order to explore the numerical convergence in the properties of the particle mid-plane layer.

We find that both the vertical and horizontal dimensions of the simulation are indeed significant factors in the nonlinear evolution of the streaming instability.  With increasing vertical domain, the particle concentrations show greater variance in their densities and migration speeds, and more stochastic events of their merging and splitting are observed.  In contrast to previous works, we begin to produce multiple, well separated, axisymmetric filamentary structures with increasing horizontal domain.  We are able to measure the typical radial separation of these structures in a sedimented particle layer driven by the streaming instability.  For particles with a Stokes number of $\sim$0.3 moving under a head wind due to the gas with an azimuthal velocity difference of $\sim$5\% local speed of sound, the mean separation of the resulting filaments is on the order of $\sim$0.2 local gas scale heights, when solid-to-gas mass ratio is $\sim$0.02.  Its possible dependence on the particle size, the radial pressure gradient of the gas, or the solid abundance remains to be investigated.  Nevertheless, this work offers the first measurement to characterize the size of the feeding zone of planetesimal formation, which may be an additional component in determining the composition of the asteroids in the Solar System.

Given that the particle layer interacts with the gas over at least one gas scale height, as shown by this work, further consideration of the streaming instability in a non-ideal magnetized disk is warranted.  In a layered-accretion disk model dominated by ohmic resistance, however quiescent of the gas in the mid-plane, there still exists non-negligible perturbations propagating down from the turbulent surface layer \citep{FS03,OMM07}.  Magneto-centrifugal winds may be launched from the surface layer when ambipolar diffusion prevails \citep{BS13}.  Hall drift could also play an important role in the dynamics of protoplanetary disks \citep{KL13,xB14,LKF14,OD14}.  How the streaming instability interacts with these non-ideal MHD effects is an important topic for future investigations.

\acknowledgments
We thank Alexander Krot for his comments on how meteorite classes may reflect the formation process for asteroids.  We thank the anonymous referee for further clarification of the manuscript.  The simulations reported in this paper were conducted on the Alarik system at LUNARC Lund University under Swedish National Infrastructure for Computing allocations SNIC001-12-148 and SNIC2013-1-205.  This research was supported by the European Research Council under ERC Starting Grant agreement 278675-PEBBLE2PLANET.  A.~J.\ is grateful for financial support from the Knut and Alice Wallenberg Foundation and from the Swedish Research Council (grant 2010-3710).


\end{CJK*}

\begin{thebibliography}{}

\bibitem[Adachi et al.(1976)]{AHN76}
Adachi, I., Hayashi, C., \& Nakazawa, K.\ 1976, Progress of Theoretical Physics, 56, 1756

\bibitem[Bai(2014)]{xB14}
Bai, X.-N.\ 2014, \apj, submitted (arXiv:1402.7102)

\bibitem[Bai \& Stone(2010a)]{BS10a}
Bai, X.-N., \& Stone, J.~M.\ 2010a, \apjl, 722, L220

\bibitem[Bai \& Stone(2010b)]{BS10b}
Bai, X.-N., \& Stone, J.~M.\ 2010b, \apj, 722, 1437

\bibitem[Bai \& Stone(2010c)]{BS10c}
Bai, X.-N., \& Stone, J.~M.\ 2010c, \apjs, 190, 297

\bibitem[Bai \& Stone(2013)]{BS13}
Bai, X.-N., \& Stone, J.~M.\ 2013, \apj, 769, 76

\bibitem[Balbus \& Hawley(1991)]{BH91}
Balbus, S.~A., \& Hawley, J.~F.\ 1991, \apj, 376, 214

\bibitem[Balsara et al.(2009)]{BT09}
Balsara, D.~S., Tilley, D.~A., Rettig, T., \& Brittain, S.~D.\ 2009, \mnras, 397, 24

\bibitem[Brandenburg \& Dobler(2002)]{BD02}
Brandenburg, A., \& Dobler, W.\ 2002, Computer Physics Communications, 147, 471

\bibitem[Brandenburg et al.(1995)]{BN95}
Brandenburg, A., Nordlund, A., Stein, R.~F., \& Torkelsson, U.\ 1995, \apj, 446, 741

\bibitem[Chiang(2008)]{eC08}
Chiang, E.\ 2008, \apj, 675, 1549

\bibitem[Davis et al.(2010)]{DSP10}
Davis, S.~W., Stone, J.~M., \& Pessah, M.~E.\ 2010, \apj, 713, 52

\bibitem[Fleming \& Stone(2003)]{FS03}
Fleming, T., \& Stone, J.~M.\ 2003, \apj, 585, 908

\bibitem[Garaud et al.(2013)]{GM13}
Garaud, P., Meru, F., Galvagni, M., \& Olczak, C.\ 2013, \apj, 764, 146

\bibitem[Goldreich \& Lynden-Bell(1965)]{GL65}
Goldreich, P., \& Lynden-Bell, D.\ 1965, \mnras, 130, 125

\bibitem[Goldreich \& Ward(1973)]{GW73}
Goldreich, P., \& Ward, W.~R.\ 1973, \apj, 183, 1051

\bibitem[Goodman \& Pindor(2000)]{GP00}
Goodman, J., \& Pindor, B.\ 2000, \icarus, 148, 537

\bibitem[Hawley et al.(1995)]{HGB95}
Hawley, J.~F., Gammie, C.~F., \& Balbus, S.~A.\ 1995, \apj, 440, 742

\bibitem[Hayashi(1981)]{cH81}
Hayashi, C.\ 1981, Progress of Theoretical Physics Supplement, 70, 35

\bibitem[Johansen et al.(2014)]{JB14}
Johansen, A., Blum, J., Tanaka, H., Ormel, C., Bizzarro, M., \& Rickman, H., Protostars and Planets VI, submitted

\bibitem[Johansen et al.(2011)]{JKH11}
Johansen, A., Klahr, H., \& Henning, T.\ 2011, \aap, 529, A62

\bibitem[Johansen et al.(2007)]{JO07}
Johansen, A., Oishi, J.~S., Mac Low, M.-M., Klahr, H., Henning, T., \& Youdin, A.\ 2007, \nat, 448, 1022

\bibitem[Johansen \& Youdin(2007)]{JY07}
Johansen, A., \& Youdin, A.\ 2007, \apj, 662, 627

\bibitem[Johansen et al.(2012)]{JYL12}
Johansen, A., Youdin, A.~N., \& Lithwick, Y.\ 2012, \aap, 537, A125

\bibitem[Johansen et al.(2009)]{JYM09}
Johansen, A., Youdin, A., \& Mac Low, M.-M.\ 2009, \apjl, 704, L75

\bibitem[Kato et al.(2012)]{KFI12}
Kato, M.~T., Fujimoto, M., \& Ida, S.\ 2012, \apj, 747, 11

\bibitem[Kowalik et al.(2013)]{KH13}
Kowalik, K., Hanasz, M., W{\'o}lta{\'n}ski, D., \& Gawryszczak, A.\ 2013, \mnras, 434, 1460

\bibitem[Kunz \& Lesur(2013)]{KL13}
Kunz, M.~W., \& Lesur, G.\ 2013, \mnras, 434, 2295

\bibitem[Lesur et al.(2014)]{LKF14}
Lesur, G., Kunz, M.~W., \& Fromang, S.\ 2014, \aap, submitted (arXiv:1402.4133)

\bibitem[Nakagawa et al.(1986)]{NSH86}
Nakagawa, Y., Sekiya, M., \& Hayashi, C.\ 1986, \icarus, 67, 375 

\bibitem[O'Keeffe \& Downes(2014)]{OD14}
O'Keeffe, W., \& Downes, T.~P.\ 2014, \mnras, accepted (arXiv:1403.8149)

\bibitem[Oishi et al.(2007)]{OMM07}
Oishi, J.~S., Mac Low, M.-M., \& Menou, K.\ 2007, \apj, 670, 805

\bibitem[Scott \& Krot(2003)]{SK03}
Scott, E.~R.~D., \& Krot, A.~N.\ 2003, Treatise on Geochemistry, 1, 143

\bibitem[Sekiya(1998)]{mS98}
Sekiya, M.\ 1998, \icarus, 133, 298

\bibitem[Weidenschilling(1977)]{sW77}
Weidenschilling, S.~J.\ 1977, \mnras, 180, 57

\bibitem[Weidenschilling(1980)]{sW80}
Weidenschilling, S.~J.\ 1980, \icarus, 44, 172

\bibitem[Windmark et al.(2012a)]{WB12a}
Windmark, F., Birnstiel, T., G{\"u}ttler, C., Blum, J., Dullemond, C. P., \& Henning, T.\ 2012a, \aap, 540, A73

\bibitem[Windmark et al.(2012b)]{WB12b}
Windmark, F., Birnstiel, T., Ormel, C.~W., \& Dullemond, C.~P.\ 2012b, \aap, 544, L16

\bibitem[Youdin \& Goodman(2005)]{YG05}
Youdin, A.~N., \& Goodman, J.\ 2005, \apj, 620, 459

\bibitem[Youdin \& Johansen(2007)]{YJ07}
Youdin, A., \& Johansen, A.\ 2007, \apj, 662, 613

\bibitem[Youdin \& Shu(2002)]{YS02}
Youdin, A.~N., \& Shu, F.~H.\ 2002, \apj, 580, 494

\bibitem[Zsom et al.(2010)]{ZO10}
Zsom, A., Ormel, C.~W., G{\"u}ttler, C., Blum, J., \& Dullemond, C.~P.\ 2010, \aap, 513, A57

\end{thebibliography}
\end{document}